\documentclass[a4paper,11pt]{article}
\usepackage{pos}
\usepackage{siunitx}
\usepackage{subcaption}
\makeatletter

\title{Progress towards stereo observation of ultra-high-energy cosmic rays with Fluorescence detector Array of Single-pixel Telescopes}
\ShortTitle{FAST Proggress for Stereo Observation}

\author*[a]{Shunsuke~Sakurai}
\author[b]{Justin~Albury}
\author[b]{Jose~Bellido}
\author[a]{Fraser~Bradfield}
\author[c]{Karel~Cerny}
\author[d]{Ladislav~Chytka}
\author[e]{John~Farmer}
\author[a]{Toshihiro~Fujii}
\author[c]{Petr~Hamal}
\author[c]{Pavel~Horvath}
\author[c]{Miroslav~Hrabovsky}
\author[c]{Vlastimil~Jilek}
\author[c,d]{Jakub~Kmec}
\author[c]{Jiri~Kvita}
\author[e]{Max~Malacari}
\author[d]{Dusan~Mandat}
\author[f]{Massimo~Mastrodicasa}
\author[g]{John~N.~Matthews}
\author[c]{Stanislav~Michal}
\author[h]{Hiromu~Nagasawa}
\author[h]{Hiroki~Namba}
\author[i]{Marcus~Niechciol}
\author[c]{Libor~Nozka}
\author[d]{Miroslav~Palatka}
\author[d]{Miroslav~Pech}
\author[e]{Paolo~Privitera}
\author[f]{Francesco~Salamida}
\author[d]{Petr~Schovanek}
\author[e]{Radomir~Smida}
\author[c]{Daniel~Stanik}
\author[c]{Zuzana~Svozilikova}
\author[j]{Akimichi~Taketa}
\author[h]{Kenta~Terauchi}
\author[g]{Stan~B.~Thomas}
\author[c,d]{Petr~Travnicek}
\author[d]{Martin~Vacula}

\small
{
\affiliation[a]{Graduate School of Science, Osaka Metropolitan University, Sumiyoshi-ku, Osaka, Japan}
\affiliation[b]{Department of Physics, University of Adelaide, Adelaide, S.A., Australia}
\affiliation[c]{Joint Laboratory of Optics of PU and IF of CAS, Palacky University, Olomouc, Czech Republic}

\affiliation[d]{Institute of Physics of the Academy of Sciences of the Czech Republic, Prague, Czech Republic}
\affiliation[e]{Kavli Institute for Cosmological Physics, University of Chicago, Chicago, IL, USA}

\affiliation[f]{Department of Physical and Chemical Sciences, University of L’Aquila and INFN LNGS}

\affiliation[g]{High Energy Astrophysics Institute and Department of Physics and Astronomy, University of Utah, Salt Lake City, UT, USA}

\affiliation[h]{Graduate School of Science, Kyoto University, Sakyo-ku, Kyoto, Japan}
\affiliation[i]{Department of Physics, University of Siegen, Germany}
\affiliation[j]{Earthquake Research Institute, University of Tokyo, Bunkyo-ku, Tokyo, Japan}
}
\emailAdd{ssakurai@omu.ac.jp}

\abstract{Ultra-high-energy cosmic rays (UHECRs) are the most energetic particles ever detected.
Cosmic rays that achieve the highest energies are rare, and their flux at Earth is extremely low.
As a result, next-generation experiments with large effective areas are required and under development.
The Fluorescence detector Array of Single-pixel Telescopes (FAST) is one such project.
Although observation time is limited compared with ground particle detectors, it enables direct measurements of $X_\mathrm{max}$, a crucial parameter sensitive to the primary cosmic-ray composition.
FAST will achieve large-area coverage by significantly reducing the cost of telescopes. This necessitates a simplified telescope compared to conventional designs.
Demonstrating the feasibility of our telescope and observational method is essential.
To validate the FAST concept, prototype telescopes have been deployed at the Pierre Auger Observatory and the Telescope Array experiment.}

\ConferenceLogo{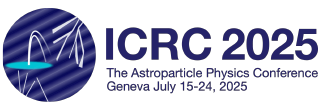}

\FullConference{39th International Cosmic Ray Conference (ICRC2025)\\
 15–24 July 2025\\
Geneva, Switzerland\\}

\begin{document}
\maketitle

\section{Ultra-high-energy comic rays}
\label{sec:intro}
Because charged particles follow helical paths in magnetic fields, their confinement scale is set by the Larmor radius $r_{\mathrm{L}}=pc/(ZeB)$; consequently, the particle energy (i.e. magnetic rigidity) encodes information on the size and field strength of cosmic accelerators. 
Lower-energy cosmic rays are expected to originate predominantly from nearby sources, whereas the highest-energy particles should arrive from increasingly distant objects.

Over the past decade, 
multi-messenger astronomy has dramatically advanced our understanding of cosmic-ray origins.
Although magnetic deflection erases directional information for the charged primaries, neutral secondaries—$\gamma$ rays and neutrinos produced in interactions of cosmic rays with ambient matter and radiation—propagate rectilinearly and thus pinpoint the acceleration sites.
In $\gamma$-ray observations, 
LHAASO has measured $\gamma$ rays up to \qty{1.4e15}{eV} from a dozen Galactic sources, establishing the existence of PeVatrons capable of accelerating particles beyond \qty{e16}{eV}. within the Milky Way as the $\gamma$-ray source~\cite{cao_ultrahigh-energy_2021, lhaaso_collaboration_ultrahigh-energy_2024}.
Neutrino astronomy has progressed apace. 
IceCube reported the \qty{6.3e15}{eV} astro-physical neutrino, which has been detected via the Glashow resonance~\cite{the_icecube_collaboration_detection_2021}.
In 2025, KM3NeT reported a candidate ultra-high-energy muon with a deposited energy of approximately \qty{1.2e17}{eV}, hinting at either extreme acceleration environments or a cosmogenic origin~\cite{the_km3net_collaboration_observation_2025}.

By contrast, direct measurements of the ultra-high-energy cosmic rays (UHECRs) have yielded unexpected results.
The Telescope Array (TA) in the Northern Hemisphere and the Pierre Auger Observatory (Auger) in the Southern Hemisphere currently provide the most precise data above \qty{e18}{eV}.
Although analyses of the energy spectrum’s directional dependence and cosmic-ray anisotropies are being actively pursued~\cite{thepierreaugercollaboration2025energyspectrumultrahighenergy,thetelescopearraycollaboration2024observationdeclinationdependencecosmic,Abbasi_2018,AbdulHalim_2025}, high-precision studies based on fine spatial subdivisions require a significantly larger number of events. 
Due to limitations in statistics, such analyses typically do not incorporate information on the depth of shower maximum ($X_{\mathrm{max}}$). 
To include $X_{\mathrm{max}}$ in these studies, large-scale observations using fluorescence telescopes represent one of the most promising approaches~\cite{ahlers_ideas_2025}.
While Auger has detected a large-scale dipole anisotropy above \qty{8e18}{eV}, no significant clustering is observed at energies exceeding \qty{e20}{eV}~\cite{doi:10.1126/science.aan4338}.
Moreover, composition studies employing sophisticated analyses, including deep-learning techniques, consistently indicate a dominance of intermediate-to-heavy nuclei at the highest energies~\cite{PhysRevLett.134.021001}.
An emblematic example is the \qty{2.44e20}{eV} “Amaterasu” event recorded by TA in 2023, whose arrival direction lies toward a region conspicuously devoid of plausible astrophysical accelerators, challenging proton-only source scenarios~\cite{doi:10.1126/science.abo5095}.

To overcome the limitations imposed by the vanishing flux ($\lesssim$ \qty{1}{km^{-2}.sr^{-1}.yr^{-1}} at \qty{e20}{eV}), next-generation observatories with enormous ground coverage are being proposed.
Among the cost-effective concepts is the Fluorescence detector Array of Single-pixel Telescopes (FAST), whose full-scale prototype has recently demonstrated the feasibility of covering thousands of square kilometres with relatively modest resources.

\section{Fluorescence detector Array of Single-pixel Telescope}
\label{sec:fast}
The Fluorescence detector Array of Single-pixel Telescopes (FAST) is a next-generation, ground-based experiment designed to collect high-statistics observations of cosmic rays with energies exceeding \qty{1e19}{eV}.
By recording atmospheric fluorescence photons, FAST reconstructs the geometry, energy, and mass composition of extensive air showers (EASs) induced from UHECRs.

Following the world’s first EAS detection using a benchtop prototype equipped with a single photomultiplier tube (PMT) in 2014~\cite{fujii_detection_2016}, we developed a simplified telescope comprising a segmented 1.6-m mirror, a UV band-pass filter, and a four-PMT focal-plane camera. 
Routine observations have been conducted ever since the first full-scale prototype was deployed~\cite{malacari_first_2020}. 
The system is now operated remotely and autonomously, primarily during moonless nights.
As of July 2025, the cumulative live times are about \qty{3500} hours for the installation at the Pierre Auger Observatory in Mendoza, Argentina (FAST@Auger), and about \qty{1000} hours for the installation at the Telescope Array site in Utah, USA (FAST@TA). 
These data are essential not only for validating the FAST concept itself, but also for enabling a rare cross-calibration between the Auger and Telescope Array experiments, as identical detectors are operated in both hemispheres.

In this proceedings contribution, we present the latest result of energy spectrum and shower-maximum depth ($X_{\mathrm{max}}$) results obtained at the two sites, together with the current status of the FAST-mini array, which is being constructed to demonstrate stereoscopic observations with FAST.

\section{Latest result with proto type telescopes}
\label{sec:mono}
Since 2016, FAST has conducted test observations using prototype telescopes~\cite{malacari_first_2020}. 
We have continued these observations to better understand the FAST telescopes and their behavior~\cite{fujii_prototype_2017,fujii_latest_2021,fujii_observing_2019,sakurai_detecting_2023}.
The FAST prototype is currently recording events triggered by neighboring fluorescence telescopes. 
Observations are ongoing in collaboration with these instruments.

\begin{figure}[ht]
 \centering
\begin{minipage}[t]{0.48\textwidth}
 \includegraphics[width=0.9\columnwidth]{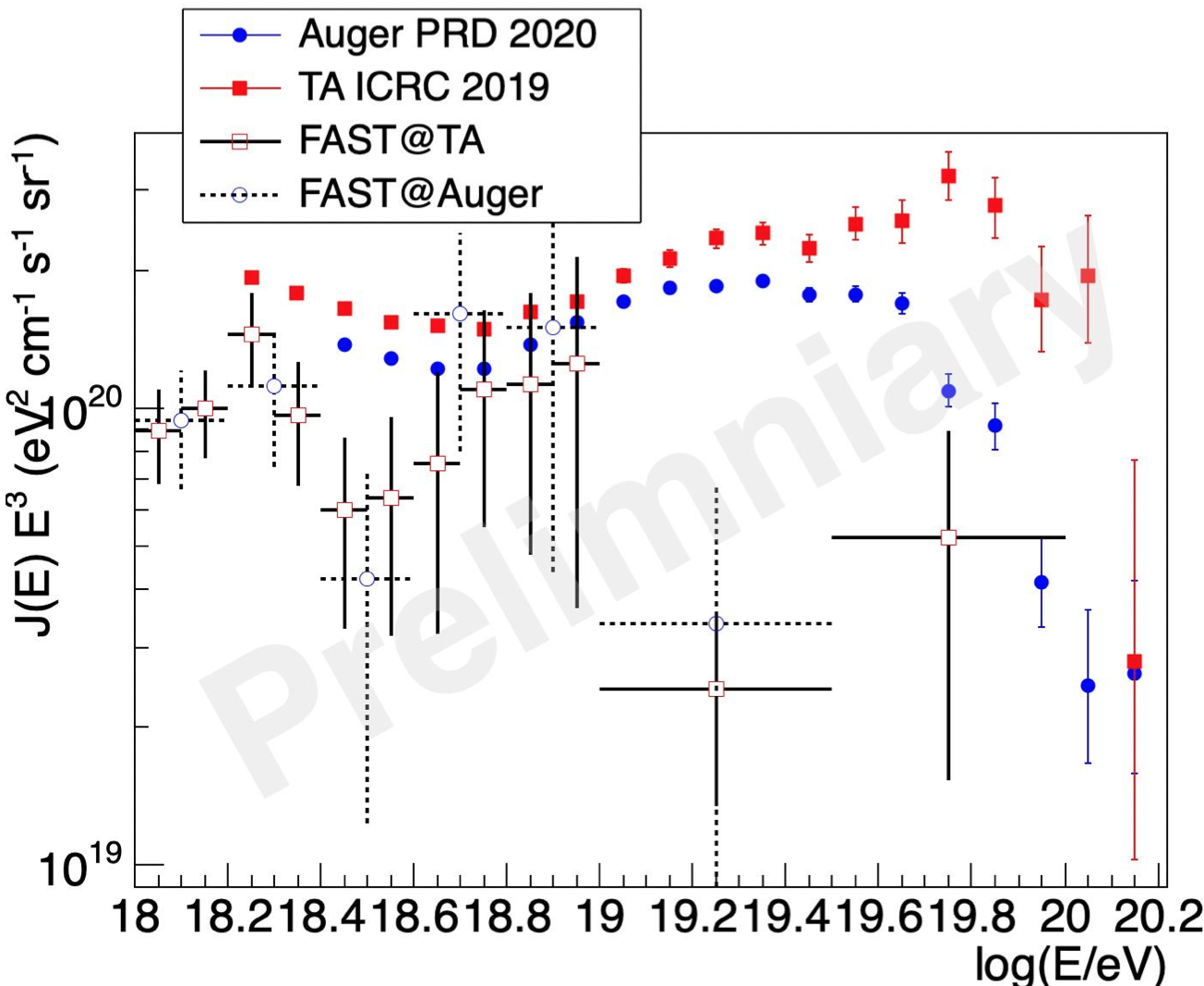}
 \subcaption{Energy spectrum}
\end{minipage}
\begin{minipage}[t]{0.48\textwidth}
 \includegraphics[width=0.9\textwidth]{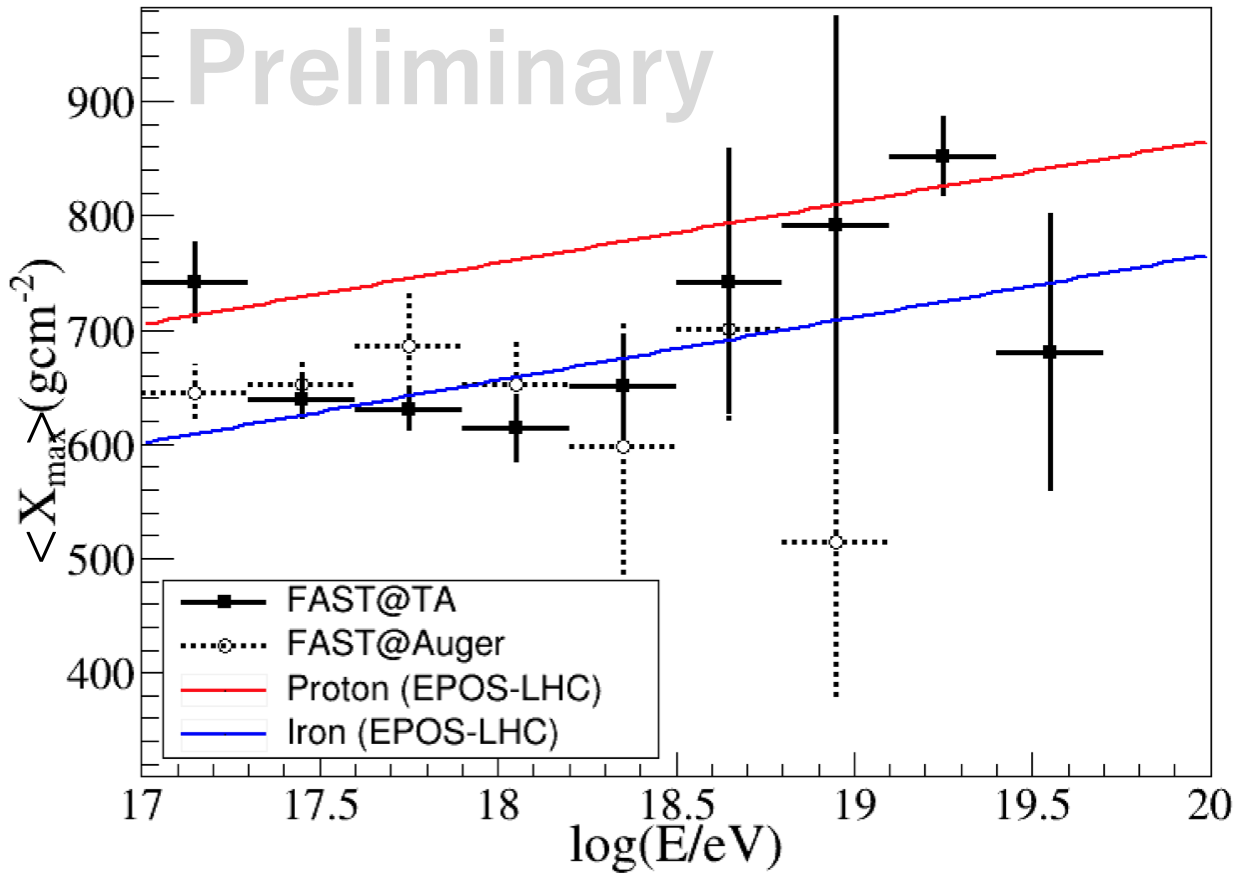}
 \subcaption{Maximum shower development height $X_\mathrm{max}$}
\end{minipage}
\caption{Latest analysis result with the FAST prototypes at the Auger and the TA site~\cite{bradfield_latest_2025}.}
\label{fig:physres}
\end{figure}

As shown in Figure \ref{fig:physres}, the reconstructed energy spectra are within an order-of-magnitude from the Auger and TA spectrum. 
Concerning the dip centered at around \qty{3e18}{eV}, we need further investigation to understand the origin.
Data corresponding to \qty{247}{hours} and \qty{122}{hours} of observation were used, respectively~\cite{bradfield_latest_2025}.
For the data–MC comparison, reconstruction parameters from Auger and TA were employed.
The distributions of both data and simulation show marginal agreement~\cite{bradfield_latest_2025}.
It should be noted that the energy spectrum was calculated using an exposure that considers only the trigger efficiency. 
We assumed that all detected events could also be successfully reconstructed by the FAST top-down reconstruction. 
A more detailed calculation of the spectrum is needed.

\begin{figure}[h]
\centering
\includegraphics[width=0.8\columnwidth]{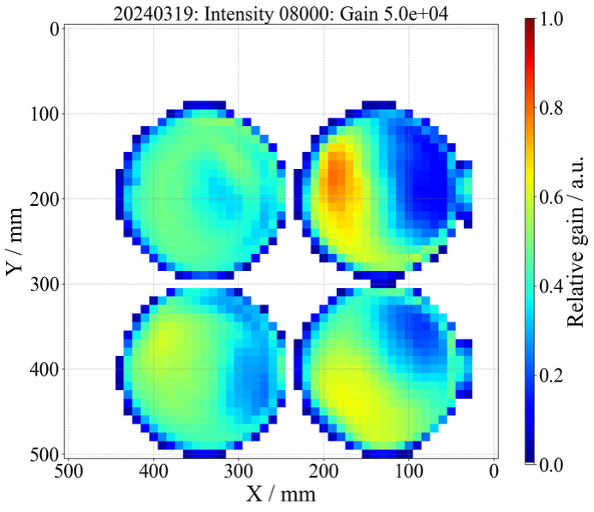}
\caption{Non uniformity structure measured on site.}
\label{fig:nonuniform}     
\end{figure}

To assess the systematic uncertainties of the telescope, we are continuing calibration work using on-site equipment.
A scan of the PMT’s relative sensitivity over its surface revealed a strong positional dependence in the current PMT camera.
Incorporating this effect into Monte Carlo simulations reduces reconstruction bias in $X_\mathrm{max}$.
Calibration measurements are ongoing to accurately characterize the FAST telescope system.
On-site measurements of PMT cathode detection efficiency (non-uniformity) were carried out.
Previous studies have shown that the structure of this non-uniformity strongly depends on the orientation with respect to the magnetic field and the rotation angle of the PMT. 
Accordingly, we measured all the PMTs used at the TA site, as shown in Figure \ref{fig:nonuniform}.

In addition, FAST recently succeeded in recording events from the EarthCARE satellite, which emits UV laser pulses to investigate clouds~\cite{AProposalofPulsePairDopplerOperationonaSpaceborneCloudProfilingRadarintheWBand}. 
These events offer a promising opportunity for cross-calibration between FAST and existing instruments.

\section{Preparation for stereo observation}
\label{sec:stereo}

Test observations conducted thus far have revealed that atmospheric fluorescence detection from a single site imposes limitations on the accuracy with which air shower parameters can be reconstructed.
Stereoscopic observation plays an important role in improving event detection and reconstruction accuracy of air showers.
Performance estimates based on stereo observations with reasonable data selection indicate approximately \qty{15}{g.cm^{-2}} for $X_\mathrm{max}$ and $E \sim$\qty{10}{\percent} resolution will be expected, respectively.

Top-down reconstruction--based on maximum likelihood searches using simulated shower profiles-- strongly depends on the initial conditions~\cite{bradfield_reconstruction_2023}.
Analysis results from Auger FD and TA FD greatly help maintain the reconstruction performance in top-down reconstruction.
However, FAST aims to operate and analyze data independently of existing instruments.
Machine learning techniques are being explored as a means to provide a good initial condition for top-down reconstruction analysis.
In addition to standard feed-forward neural networks, Long Short-Term Memory (LSTM) networks have also been tested as possible architectures~\cite{10.1162/neco.1997.9.8.1735}.
The TESFEL library was also investigated for preprocessing signal waveforms to serve as neural network inputs~\cite{barandas2020tsfel}.

\begin{figure}[h]
\centering
\includegraphics[width=0.8\columnwidth]{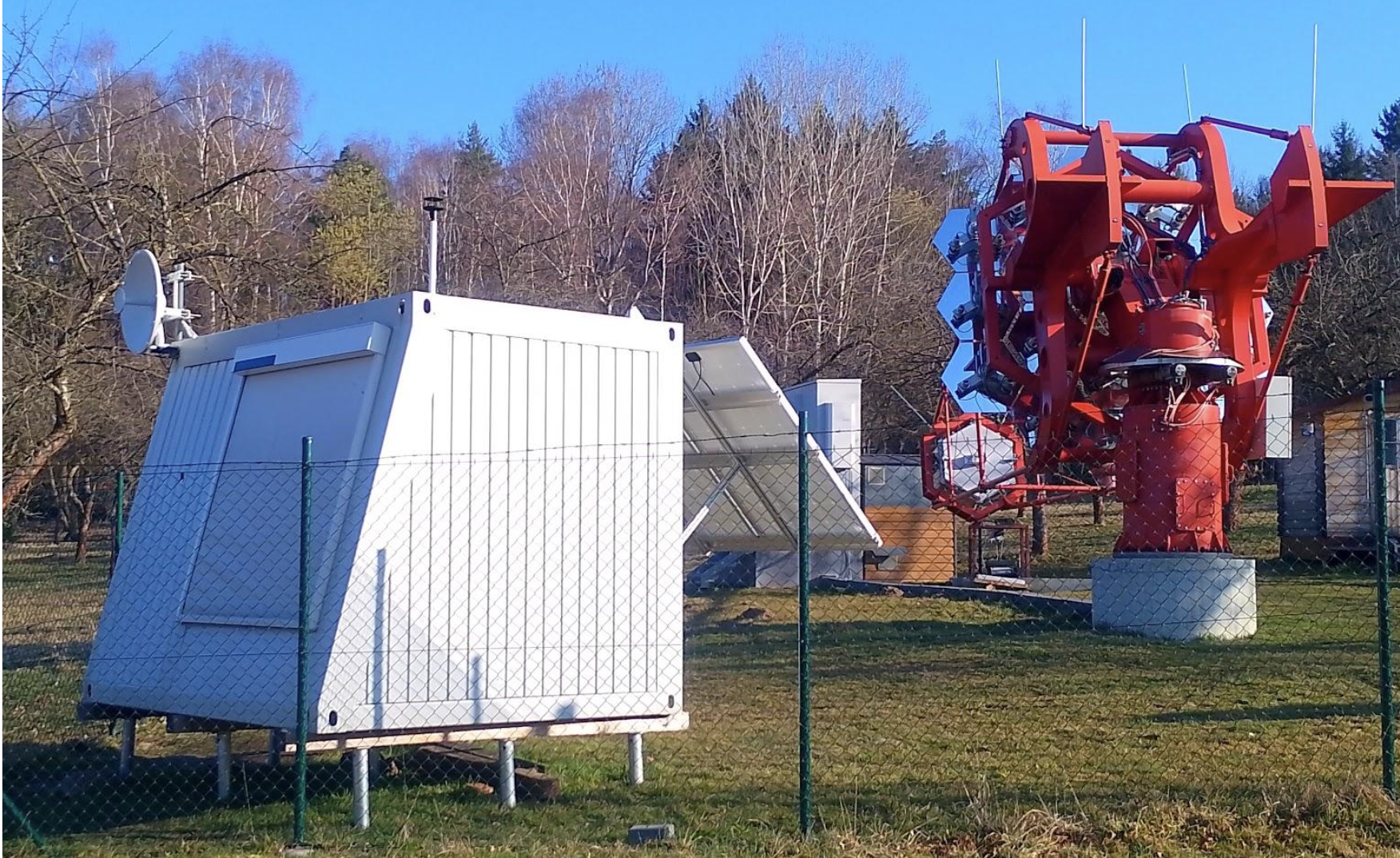}
\caption{The new-design FAST telescope being tested at Ondrejov observatory.}
\label{fig:fast3}     
\end{figure}

Preparation of two additional two telescopes is being prepared and developed.
They are being built with an upgrade design that enables stand-alone operation.
Electronics such as high-voltage module and amplifiers were newly developed for these stand-alone telescopes, with an emphasis on low power consumption. 
To asses observational capability, sensors for temperature, humidity, rain, and wind have been installed and their functionality confirmed.
Figure \ref{fig:fast3} shows the new prototype (FAST-Field) being tested at Ondrejov observatory, Czech Republic.

New mirrors have been produced for the new telescopes.
Grounding and polishing were omitted from the production process while preserving sufficient focusing performance.
Moreover, the average reflectivity for wavelength below \qty{400}{nm} has improved by approximately \qty{4}{\percent} in average. 
The new hut design, including the telescope structure, facilitates transport and deployment at observation sites easily.
A newly installed curtain behind the UV filters blocks sunlight in place of rolling shutters for emergency cases.
Solar panels have also been installed to charge butteries during the daytime,
enabling nighttime operation using charged power, which also supports slow control systems.
To enable the long-distance communication, a \qty{5}{GHz}, WiFi antenna has been installed on the top of hut.
Currently, Wi-Fi is planed to use exclusively for data transfer and slow control, but inter-telescope triggering is also need to be tested in future.

In 2025, additional two telescope will be installed \qty{11}{km} away from the Los Leones (LL) at Auger site.
Those telescopes will point toward LL, together with existing prototypes, will form a small array.
Our focus will be lowering the detection energy threshold to increase the number of "good" stereo events.
In the coming years, two more telescopes will be added to form a triangular array with spacing \qty{16}{km} and test a more realistic telescope configuration.
We expect roughly \num{100} events per year with stereo conditions.

\section{Summary and Prospects}
\label{sec:summary}

As discussed in Section \ref{sec:stereo}, stereo observation is capable for detecting UHECRs with reasonable performance.
A low-cost fluorescence detector, such as FAST, can contribute not only in the UHECR observation, but also wider range of cosmic ray experiments.
Validating stereo observation method and data availability shed light for the various application of the simplified design of the telescope.

Latest performance estimation showed a promising performance as \qty{10}{\gram\per\cm}.
We will install stereo array at Pierre Auger Observatory and will extend the array to realistic situation.
Telescope installation is planed in this and coming years.
Combining the data from the surface detector array with simple Fluorescence detector would be worth to try it.

\section*{Acknowledgements}
This work was supported by JSPS KAKENHI Grant Number 25H00647, 21H04470. This work was partially carried out by the joint research program of the Institute for Cosmic Ray Research (ICRR) at the University of Tokyo. This work was supported in part by NSF grant PHY-1713764, PHY-1412261 and by the Kavli Institute for Cosmological Physics at the University of Chicago through grant NSF PHY-1125897 and an endowment from the Kavli Foundation and its founder Fred Kavli. The Czech authors gratefully acknowledge the support of the Ministry of Education, Youth and Sports of the Czech Republic project No. CZ.02.1.01/0.0/17\_049/0008422, CZ.02.01.01/00/22\_008/0004632, LM2023032, Czech Science Foundation project GACR 23-07110S and the support of the Czech Academy of Sciences and Japan Society for the Promotion of Science within the bilateral joint research project with Osaka Metropolitan University (Mobility Plus project JSPS 21-10). The authors thank the Pierre Auger and Telescope Array Collaborations for providing logistic support and part of the instrumentation to perform the FAST prototype measurement and for productive discussions.

\bibliographystyle{JHEP}
\bibliography{skeleton}

\end{document}